# Neuromorphic Spintronics

J. Grollier,[1*] D. Querlioz,[2] K. Y. Camsari,[3] K. Everschor-Sitte,[4] S. Fukami,[5] M. D. Stiles[6]

[1]Unité Mixte de Physique CNRS, Thales, Univ. Paris-Sud, Université Paris-Saclay, 91767 Palaiseau, France
[2]Centre de Nanosciences et de Nanotechnologies, Univ. Paris-Sud, CNRS, Université Paris-Saclay, 91405 Orsay, France
[3]School of Electrical & Computer Engineering, Purdue University, West Lafayette, Indiana 47907 USA
[4]Institute of Physics, Johannes Gutenberg University Mainz, D-55099 Mainz, Germany
[5] Research Institute of Electrical Communication, Tohoku University, Sendai, Miyagi 9808577, Japan
[6]National Institute of Standards and Technology, Gaithersburg, Maryland 20899, USA

* julie.grollier@cnrs-thales.fr

**Neuromorphic computing uses basic principles inspired by the brain to design circuits that perform artificial intelligence tasks with superior energy efficiency. Traditional approaches have been limited by the energy area of artificial neurons and synapses realized with conventional electronic devices. In recent years, multiple groups have demonstrated that spintronic nanodevices, which exploit the magnetic as well as electrical properties of electrons, can increase the energy efficiency and decrease the area of these circuits. Among the variety of spintronic devices that have been used, magnetic tunnel junctions play a prominent role because of their established compatibility with standard integrated circuits and their multifunctionality. Magnetic tunnel junctions can serve as synapses, storing connection weights, functioning as local, nonvolatile digital memory or as continuously varying resistances. As nano-oscillators, they can serve as neurons, emulating the oscillatory behavior of sets of biological neurons. As superparamagnets, they can do so by emulating the random spiking of biological neurons. Magnetic textures like domain walls or skyrmions can be configured to function as neurons through their non-linear dynamics. Several implementations of neuromorphic computing with spintronic devices demonstrate their promise in this context. Used as variable resistance synapses, magnetic tunnel junctions perform pattern recognition in an associative memory. As oscillators, they perform spoken digit recognition in reservoir computing and when coupled together, classification of signals. As superparamagnets, they perform population coding and probabilistic computing. Simulations demonstrate that arrays of nanomagnets and films of skyrmions can operate as components of neuromorphic computers. While these examples show the unique promise of spintronics in this field, there are several challenges to scaling up, including the efficiency of coupling between devices and the relatively low ratio of maximum to minimum resistances in the individual devices.**

## I- Neuromorphic computing is the path to low energy Artificial Intelligence

Artificial Intelligence has experienced unprecedented progress in recent years, promising to transform multiples areas of how we live and how we work. However, this development comes with a considerable challenge: the energy consumption associated with existing approaches[1], making it imperative that we devise ways to process data more efficiently. One approach is to emulate the brain's processing, which is much more efficient than current processors at cognitive tasks like image and speech recognition. Although modern Artificial Intelligence relies on algorithms known as deep neural networks, their operation on processors radically differs from the brain. Modern computers and graphics cards have been designed to solve complicated numerical problems with high precision, while the brain uses many parallel low precision calculations to, for example, recognize a face. Computers achieve high precision using digital information encoding but the brain achieves its energy efficiency with lower precision analog encoding. Modern computers consume substantial energy shuttling information between storage and the processor, while the brain stores information locally where it is processed.

The elemental devices in the brain and in modern computers play different roles. Modern computers use transistors that are voltage-controlled switches and cannot provide memory in a compact form. The brain has two primary elemental units, synapses and neurons. In their simplest abstraction, synapses connect neurons with a connection strength, called a weight, which provides the memory function. Neurons receive inputs from many other neurons, integrate those responses, and emit spikes, called action potentials, which provide the input for subsequent neurons. Emulating the organization of the brain by using transistors to function like neurons and synapses requires many transistors[2], using more energy and requiring greater area (typically hundreds to thousands of square micrometers[3]) than appropriate for many modern embedded applications.

The research reviewed in this article attempts to develop compact and low power computational systems using spintronic devices as an alternative to the large number of transistors needed to emulate the functions of neurons and synapses and connect those functional blocks together[4,5]. At the device level, the emphasis is on magnetic tunnel junctions (see Fig. 1a), which are being developed for non-volatile memory, (see Fig. 1b) in the back-end-of-line of Complementary Metal Oxide Semiconductor (CMOS) chips[6]. Major commercial foundries have now incorporated these devices in their processes[7]. This compatibility and the variety of functionalities available by changing geometries make magnetic tunnel junctions attractive candidates for efficient computing.

Magnetic tunnel junctions have several features that other technologies[8], both existing and emerging, do not combine; *e.g.*, nonvolatility, outstanding read/write endurance, high-speed and CMOS-compatible-voltage operation capability, high scalability, and back-end-of-line compatibility. However, the ratio of their maximum to minimum conductance (ON/OFF ratio) is typically around three whereas it can reach thousands for other resistive switching memories[9].

Spintronic approaches extend beyond the use of magnetic tunnel junctions as binary memory cells. An advantage of spintronics for neuromorphic computing is the multifunctionality that it offers, allowing designers to craft behaviors ranging from non-volatile through plastic, oscillatory, to stochastic, all from very similar materials. This enables the design of diverse building blocks mimicking key features of biological synapses and neurons. In addition, spintronics enables interconnecting these building blocks without relying on just CMOS connections. Spintronic components can carry information to distant places through spin currents, microwave signals, magnetic waves, and isolated magnetic textures that can then be moved around. This multifunctionality opens a wealth of possibilities to build spintronics-based neuromorphic chips that take advantage of these additional features and communications channels, thereby decreasing the CMOS overhead where it is inefficient. Here, we review the first steps in this direction. We first describe spintronic neuromorphic building blocks and then discuss demonstrations of spintronic neuromorphic computing in small hardware systems. Finally, we analyze the advantages and disadvantages of spintronics for building larger systems.

## II- Spintronic synapses

### a. Embedding memory in the processor

In current computers, synaptic weights are stored as digitally-coded numbers in memory blocks separated from the circuits that process them. State-of-the-art neural networks can use more than a hundred million of these weights. Each time a neural network infers or learns, all these parameters must be fetched from memory for processing. Shuttling such quantities of data back and forth between memory and processing requires inordinate amounts of energy. The most straightforward way spintronics can enhance neuromorphic computing is by locating fast, non-volatile binary memory blocks very close to the processing units taking advantage of the ability to embed magnetic tunnel junctions within CMOS circuits[10]. These embedded devices also offer the possibility of turning off unused memory circuits without losing memorized information[11]. Local memory as well as energy management can be harnessed to realize high-performance energy-efficient neuromorphic chips.

Magnetic tunnel junction memory cells have been used recently to store the synaptic weights of hardware neural networks called associative memories (see Fig. 1c). Jarollahi et al.[10] have fabricated a content-driven search engine using a magnetic tunnel junction-based logic-in-memory architecture. They reduced memory needs by a factor of 13.6 and energy consumption by 89 % compared with a non-neural hardware-based search architecture using content-addressable memories. The number of clock cycles in performing search operations of the developed chip was reduced by a factor of 8.6 compared with common content addressable memories and by a factor of five orders of magnitude compared with a search engine based on a traditional processor.

Ma et al.[11] fabricated an associative processor that comprises a four-transistors and two-magnetic tunnel junctions (4T-2MTJ) spin-transfer torque magnetoresistive random-access memory. They drastically reduced the energy consumption with an intelligent powering strategy, in which only currently accessed memory cells are autonomously activated. This approach reduces power consumption by 91.2 % compared with a twin chip designed with six-transistors static random-access memory, and by more than 88.0 % compared with the latest associative memories[11]. These results show the improvements that digital magnetic tunnel junction devices can bring to neuromorphic chips. Given that spin-transfer torque magneto-resistive random-access memory is about to hit the mass market, the very first contributions of spintronics to commercial neuromorphic chips will likely rely on the use of digital magnetic memories embedded close to CMOS circuitry for low-power cognitive computing.

### b. Exploiting the inherent stochastic switching in binary magnetic tunnel junctions

An important challenge for magnetic tunnel junctions is that they are inherently prone to bit errors due to the role thermal activation plays in their switching dynamics. For conventional applications, microelectronics designers alleviate this partial unreliability with engineering solutions such as the use of junctions with a high energy barrier compared to thermal energy (leading to high programming currents), error correcting codes or specific write strategies that check the results of write operations[12]. Such solutions downgrade the energy efficiency of the devices. However, synapses, which implement the long term memory of the brain, are far from perfectly reliable[13]. Correspondingly, when magnetic tunnel junctions are used as the memory for neural networks, they do not necessarily need to have the reliability required for usual computing: neural networks are inherently resilient to bit errors. It is possible to design neural networks with synapses that have a relatively high error rate, without endangering the functionality of the whole network[14,15].

Programming errors might even be exploited when training a neural network[16,17]. Training requires repeated adjustment of the synaptic weights, usually by small amounts. One alternative approach is to make larger changes, but with reduced probabilities. This approach has little value in conventional systems, as implementing probabilities requires generating random numbers, which is energy-intensive in conventional electronics. However, magnetic tunnel junctions operating in a regime with high bit error rates, can efficiently realize this alternative approach. Vincent et al.[18] simulated stochastic switching of binary magnetic tunnel junctions and showed that they can be harnessed to implement spike-timing-dependent plasticity, a biologically inspired learning rule. Using an accurate physical model, they demonstrated unsupervised recognition of patterns in video streams. This approach reduces the memory footprint of neural networks: fewer bits are required for weights than for conventional training. For some tasks, a single bit per synapse may suffice[18]. More fundamentally, the junctions can be programed with short, low current pulses thus strongly decreasing energy consumption during learning. Embracing bit errors exploits the true energy efficiency of spin torque, whereas fighting them costs substantial energy.

### c. Spintronic memristors

We have focused until now on the use of binary magnetic tunnel junctions for neuromorphic computing, based on the natural encoding of binary information in magnetic materials through the direction of their magnetization, which points either up or down in the new generation of memories. However, synaptic weights in neural networks, as with synapses in the brain, are typically real-valued, not binary. This means that many binary magnetic tunnel junctions are needed to store a single weight, costing area and read/write energy. There is therefore a strong interest to develop analog storage elements that individually emulate synapses in neuromorphic networks. In addition to being analog and non-volatile, these components should be plastic, meaning that the long-term properties of the device can be modified by its inputs, allowing stored memories to be tweaked.

Analog, nonvolatile, and plastic resistors, now often referred to as memristors, were introduced as early as the 60's by Widrow and Hoff,[19] who used them as hardware synapses. These components have been then theorized as fundamental circuit elements by Chua in the seventies[20] and revisited experimentally in 2008 by Strukov et al.[21] with Pt-$TiO_{2-x}$-Pt nanodevices. Since then, various material systems have been used in memristive devices[9]. Memristors are particularly suited for imitating synapses. Just as synapses are non-volatile analog valves for information in the brain, memristors are non-volatile, analog valves for electrical currents. In neural networks, memristors naturally implement another important function more efficiently than CMOS circuits: the weighted sum of neural outputs by synapses. The current flowing through memristors electrically connected in parallel is the weighted sum of the memristor conductances times the input voltage[22].

Magnetic devices can function as memristive devices by storing analog information in magnetic textures[23]. For example, Wang et al. proposed a spintronic memristor[24] based on the displacement of a magnetic domain wall[25] in a spin-valve (see Fig. 2a), giving rise to lower or higher resistance states depending on the domain wall position[26]. Chanthbouala et al.[27] and Lequeux et al.[28] experimentally demonstrated this memristive functionality through domain wall motion in magnetic tunnel junctions. Huang et al.[29] simulated another concept for a spintronic memristor, based on representing analog information in the number of magnetic skyrmions (see Fig. 2b). Wadley et al. demonstrated analog-like operation in antiferromagnetic CuMnAs spintronic devices, using current-induced control of the Néel vector in submicron-scale antiferromagnetic domains[30,31]. Fukami et al. used spin-orbit torque switching to control a memristive element[32–34] in an antiferromagnet/ferromagnet bilayer system[35] (see Fig. 2c). The memristive behavior comes from the variation in the switching currents among the small magnetic domains that have varying exchange-bias magnitudes and directions at the antiferromagnet/ferromagnet interface[36].

Spintronic memristors enjoy most of the advantages of spintronic digital memory devices, making them unique building blocks for neuromorphic computing with artificial synapses. Their nonvolatility allows them to capture simultaneously the two key features that synapses need to exhibit for neuromorphic computing: learning and memory. Moreover, the high endurance of spintronic memristors allows an outstanding number of learning cycles. This feature is particularly important for adaptive applications, especially in Internet-of-Things systems. One of the biggest challenges for the spintronic memristors is scalability, *i.e.*, maintaining the analog behavior with reduced device dimensions. Overcoming this challenge requires engineering materials that are capable of hosting more magnetic domains or skyrmions in nanoscale devices.

### III-    Spintronic neurons

Until recently, the majority of the effort to use nanotechnology in hardware neural networks has focused on synapses. As synapses are much more numerous than neurons in most systems, the benefits of implementing them at the nanoscale seems more evident. In addition, neural operations in state-of-the-art deep networks are simple non-linear functions that could be implemented piecewise with a few transistors. Nevertheless, neurons in the brain have much more complicated features. They are not static objects, but excitable cells, that leakily integrate the electrical spikes that they receive from other neurons and emit a spike when their membrane potential is charged above a threshold. After firing, the membrane potential falls back to the resting state and undergoes a refractory period. A neuron receiving a constant rate of input spikes therefore fires periodically, which explains why a whole branch of computational neuroscience uses non-linear dynamics to model neurons as non-linear oscillators coupled by synapses[37–39].

When noise is high, which is often the case in biological neuron recordings, the emitted spike trains may become seemingly random. For this reason, several neuroscience approaches treat neural firing as a Poisson process and neural operations as stochastic processes[40]. These models and approaches are interesting for neuromorphic computing as they can potentially give additional features (e.g. time-dependent processing of input fluxes) or benefits (lower energy consumption by harnessing thermal processes). Spintronics, which allows the implementation of non-linear magnetization dynamics and stochastic processes at the nanoscale, gives numerous opportunities in this field[5].

#### a.    Spin-torque nano-oscillators

Spin-torque nano-oscillators (see Fig. 3a) are specific types of magnetic tunnel junctions, which can be driven into spontaneous microwave oscillations by an injected direct current[41,42]. Spin-torque nano-oscillators possess several distinctive features that are appealing for neuromorphic computing[4]. The oscillation amplitudes have memory due to finite magnetization relaxation, which can imitate the leaky integration of neurons[43,44]. They are stable and persistent, with limited drift in the behavior of their precession. The frequency and amplitude of voltage oscillations are highly non-linear as a function of current or applied field, allowing direct implementation of non-linear activation functions. In addition, their high tunability facilitates synchronization with other oscillators[45]. They can couple to other spin-torque nano-oscillators through direct exchange interactions[46–48], magnetic fields[49,50], or oscillating electrical currents due to the giant or tunneling magnetoresistance[51]. This ability to couple enables coupling many devices together through these physical interactions[52,53] to emulate the synchronization of neurons and collections of neurons in the brain to improve information sharing and processing[54].

Torrejon et al. demonstrated neuromorphic computing with a single spin-torque nano-oscillator[55] emulating a full neural network of 400 neurons using time-multiplexing[56] (see Fig. 3b). The single oscillator emulates 400 neurons by periodically devoting an interval in time for the state of each neuron and using the finite relaxation time to emulate coupling between neurons. The authors used the oscillator to implement a reservoir computer, a type of neural network especially adapted to dynamical situations[57]. The time-multiplexed nano-oscillator recognizes spoken digits from the NIST TI-46 database[58] with a precision up to 99.6 %, which is as good as is done with both much larger neurons and software simulations. The authors show that this high performance of spin-torque nano-oscillators used as neurons comes from their stability, low noise and high non-linearity.

### b. Superparamagnetic tunnel junctions

Studies of the brain suggest additional approaches for using magnetic tunnel junctions as neurons. Many experimental and theoretical works in neuroscience indicate that synapses and neurons in the brain are at least partly stochastic[59]. Some parts of the brain seem to trade reliability for energy effficieny[13,60]. Biological neurons are sometimes modeled as Poisson neurons with random spiking[61]. Since magnetic tunnel junctions are prone to stochastic effects, one can implement low energy artificial neurons by exacerbating stochastic effects, by using binary superparmagnetic tunnel junctions (see Fig. 4a). In such junctions, the energy barrier between the parallel and anti-parallel states is comparable to the thermal energy, so that even in the absence of electrical current and magnetic field, switching is triggered by thermal fluctuations.

Superparmagnetic tunnel junctions have distinctive features. First, they can be used to generate random bits simply by reading the state of the junction, an extremely low energy operation[62]. Second, they are reminiscent of Poisson neurons, with the difference that the output of such junctions is a telegraph signal whereas the output of such neurons is a spike train. The switching rate of these junctions can be controlled through spin-torques and magnetic fields[63], and used for neuromorphic computing. For example, Mizrahi et al. showed that superparamagnetic junctions can phase lock to periodic inputs[64] just like neurons in the brain, providing a mechanism for neuroscience-inspired forms of computation.

A third way to compute with superparamagnetic tunnel junctions is to use their average state rather than their transition rate. Digital electronics is based on deterministic bits that represent zero or one. Bits realized by modern CMOS transistors are used by very large-scale circuits to implement complex functions. On the other extreme, quantum computing relies on qubits, a coherent superposition of zero and one. In between these extremes, it is possible to envision probabilistic bits, or p-bits (see Fig. 4b) classical entities that fluctuate between zero and one in the presence of thermal noise[65]. Magnetic tunnel junctions with low barrier nanomagnets naturally function as a compact hardware realization of a three-terminal p-bit, allowing them to be interconnected as correlated circuits. Two possibilities to construct p-bits with magnetic tunnel junctions have been discussed, one using spin-orbit-torque for switching[65] and one using spin-transfer-torque for switching[66]. Both involve replacing the thermally stable free layers of the tunnel junctions with unstable nanomagnets, either by reducing the anisotropy or by reducing the total magnetic moment[62,67–69]. While p-bits can be implemented using CMOS circuits, implementations based on nanodevices like magnetic tunnel junctions may enable ultra-low power stochastic computing reminiscent of brain processes.

### c. Domain-wall and skyrmion based neurons

Spin-torque nano-oscillators and superparamagnetic neurons rely on magnetic tunnel junction technology. Alternative types of neurons based on magnetic solitons can also be envisioned as proposed by Sharad et al in Ref. [23]. Magnetic solitons such as domain walls and skyrmions (see Fig. 5a) can be manipulated and moved over large distances with spin-torques and spin-orbit torques[70–72]. These objects are possible vectors of information that can be used for computing. For instance, magnetic domain-wall-based logic has been studied extensively, and the basic operations that have been demonstrated can be used for neuromorphic computing[73,74]. In this context, it is possible to take advantage of the fundamentally stochastic nature of the depinning and motion of magnetic nanotextures[75–77]. In particular, the particle-like behavior of skyrmions and their thermal Brownian motion has strong analogies with neurotransmitter diffusion[78]. Simulations show that switching after cumulative domain wall motion[23], or skyrmion accumulation in a chamber[79,80,77] are spintronic analogs of leaky integrate and fire neurons.

Non-linear resistance changes in magnetic skyrmion systems[81] can be exploited for unconventional computing[82–84]. Such changes originate from an interplay of magnetoresistance effects (like the anisotropic magnetoresistance or non-collinear magnetoresistance[85,86]) combined with spin-(orbit)-torques on the skyrmions that either move or distort them. Prychynenko et al.[82] analyzed the single skyrmion resistance response based on the interplay of spin-transfer torques and the anisotropic magnetoresistance using micromagnetic calculations that self-consistently solve for the magnetization dynamics and the current path[87]. The output voltage of such a device can be converted into a synaptic current.

### IV- Neuromorphic computing with small spintronic systems

Using spintronics for neuromorphics is interesting for more than just single devices. Spin currents, spin waves or microwave emissions can be harnessed to propagate information between devices. However, assembling spintronic neurons and synapses directly in systems comes with specific challenges: controlling their coupling, and dealing with inevitable device variability. In recent years, highly promising research has started to address these points.

### a. Computing with spintronic memristors

We have seen that spintronic memristors can be used as artificial synapses. The ability to update their states given new information, that is to learn, is a key capability of artificial synapses in artificial neural networks. The state of each synaptic device is tuned by training so that the network collectively stores the information. As is discussed in other articles in this series, pattern classification has been demonstrated in perceptron networks with artificial synapses made of metal-oxide resistive devices[88] and phase-change material devices[89].

Borders et al. demonstrated a proof-of-concept associative memory (see Fig. 2d) based on an artificial neural network with spintronic synapses[90]. They employed antiferromagnet/ferromagnet spin-orbit torque switching devices with memristive functionality as described earlier[35]. The Hopfield model[91], which was originally developed from an analogy with spin glass systems, is used for memorization and association of patterns. In this model, each neuron is connected to all other neurons via synapses with variable synaptic weight and the synaptic weight matrix encodes the stored information.

To demonstrate pattern association, Borders et al. used three kinds of 3×3 block patterns, corresponding to 9-neuron systems. In this case, the synaptic weight matrix requires 36 synaptic devices due to the symmetry of the matrix. The authors constructed a Hopfield network consisting of 36 spin-orbit-torque-based memristive devices, driven by field-programmable gate arrays that emulate neurons. The system is controlled by software running on a computer. To initialize the system, electric currents corresponding to the ideal synaptic weights calculated for the three patterns based on the Hopfield model are applied to the prepared synaptic devices. Due to insufficient linearity and uniformity of the devices, the network does not remember the given patterns at this stage, requiring a learning process, based on the Hebbian learning rule[92], to compensate for the imperfection of the synaptic devices. The learning process converges with at most 20 iterations, after which the network remembers the given patterns. Importantly, this work demonstrates learning using spintronic synapses. As spintronic synapses have high endurance, neuromorphic hardware with spintronic synapses can deliver superior adaptivity through learning.

### b. Computing with synchronized spin-torque nano-oscillators

In the system that we just described, spintronic synapses were combined with conventional electronics to enable learning. Spintronic neurons can also be trained to compute. Romera et al. demonstrated classification of signals at microwave frequencies through the synchronization of spin-torque nano-oscillators[93] (see Fig. 3c). They implemented a small neural network with two layers. This network features two independent neurons in the first layer (*A* and *B*), implemented by two microwave sources delivering sinusoidal waveforms of frequency $f_A$ and $f_B$, and four all-to-all connected neurons in the second layer (labeled *i*), implemented by four spin-torque nano-oscillators that are globally coupled through long range electrical microwave connections. The microwave outputs of the first layer are sent through a stripline above the four oscillators in the second layer: the resulting microwave fields modify the oscillator dynamics. The principle of the computation is that the synchronization of two oscillators models a strong synaptic coupling between them[94]. If neuron *i* in the second layer synchronizes with neuron *A* in the first layer, the equality of their frequencies models a strong synaptic coupling. On the other hand, neuron *A* and neuron *i* having independent dynamics and frequencies models weak synaptic coupling between them. These synaptic strengths can be tuned by changing the free-running frequency of each oscillator in the second layer through the four injected direct currents that feed them. If the frequency of neuron *i* is closer to the frequency of neuron *A*, it will be more likely to synchronize with neuron *A*, corresponding to a stronger synapse.

With this approach, Romera et al.[93] trained a neural network of four coupled spin-torque nano-oscillators to classify seven American vowels (https://youtu.be/IHYnh0oJgOA). Training requires less than hundred iterations. The experimental recognition rate after training is 89 % on the test data (84 % after cross validation). This performance is significantly better than that of a multilayer perceptron trained on the same task with a similar number of parameters. In perceptrons, neurons are indeed not connected within a layer but here, the coupled oscillators interact to recognize the vowels. This result demonstrates that the dynamical properties of spin-torque nano-oscillators can be tuned to learn and that their coupling and synchronization can be harnessed to classify. The authors also showed that with this scheme, scaled-down oscillators based on state-of-the-art magnetic tunnel junctions compute with slightly lower energy consumption than optimized CMOS circuits. Developing large scale networks based on this approach requires designing arrays with hundreds of spin-torque oscillators with different frequencies but similar synchronization ranges. In addition, the simple learning rule developed in this demonstration might not easily extend to training deep networks[95]. Finding ways to tune the coupling between oscillators instead of changing their individual frequencies will be key to extend synchronization-based approaches to multilayer spintronic neural networks[96].

### c. Computing with superparamagnetic magnetic tunnel junctions

Just as the deterministic oscillations of spin-torque nano-oscillators can emulate neuron responses, the temperature-driven random fluctuations in superparamagnetic magnetic tunnel junctions can be used to imitate neural Poisson spiking dynamics. The analogous behavior of neurons and spin torque nano-oscillators can be pushed ever further. When subjected to an electrical current and the resulting spin torque, the mean frequency of superparamagnetic tunnel junctions has a bell-shaped response as a function of current (see Fig. 4c). This response is reminiscent of the stimuli-induced response of sensory neurons, such as those connected to our retina. Neuroscience has investigated how the brain relies on such curves to compute, through the paradigm of population coding, where each neuron responds with a bell curve, but each with a different mean value[61]. Through combined experiments and simulations, Mizrahi et al.[97] showed that assemblies of superparamagnetic tunnel junctions can implement neural population coding and perform complex cascaded non-linear operations on their inputs, Fig. 4(d), – the basics of deep learning. The authors illustrate how a robot equipped with such a superparamagnetic neural network could reliably learn to grasp a ball, despite component unreliability. The resilience to device unreliability is a natural benefit of population coding, as the use of a population of devices to code one real value provides a form of intrinsic error correction[98].

Additionally, Mizrahi et al.[97] designed a full combined CMOS-spintronic circuit connecting the junctions for this application. They find that a system with 128 inputs and 128 outputs consumes 23 nJ per operation during the learning phase, and 7.4 nJ when learning is finished, compared to 330 nJ per operation for an implementation based on low-power spiking CMOS neurons. The roots of this energy efficiency are threefold and highlight generic advantages of spintronics for neuromorphic computing. First, the design closely integrates sensing, memory and logic, taking advantage of the ability to integrate spintronics with CMOS. Second, the system is stochastic and computes approximately, harnessing the randomness of spintronics in a way that is more energy efficient than traditional precise electronics. Finally, the superparamagnetic tunnel junctions convert between analog (input current) and digital (spikes) information with more energy efficiency than traditional analog-to-digital conversions.

Another way to compute with superparamagnetic tunnel junctions is to solve different classes of complex problems by encoding their solutions as low-energy states of probabilistic p-bit based circuits[99]. Such circuits (see Fig. 4e) based on superparamagnetic tunnel junctions with very low barriers ($E_B \approx k_B T$) can stochastically search the vast phase-space of hard problems at high speed (from megahertz to gigahertz) in massively parallel, asynchronous networks[99]. Applications broadly relevant for two disjoint areas of research, namely machine learning and quantum computing could be targeted by such p-circuits. In the context of machine learning, the p-bit can be imagined as a hardware representation of a binary stochastic neuron[100,101], commonly used as a building block for stochastic artificial neural networks, such as Boltzmann Machines[92]. Hardware p-circuits can not only help enable low-power stochastic inference networks[102] but also accelerate learning algorithms that require repeated evaluations of correlations between interconnected binary stochastic neurons.

Quantum annealers[103,104] explore a large phase space through quantum fluctuations to address computationally hard optimization problems such as the NP-complete Traveling Salesman Problem or Integer Factorization. Simulations of networks of p-bits show that such optimization problems can also be addressed by classical p-bits[99]. For example, classical annealing using hardware p-circuits can be performed by guiding the network to energy minima. An unconventional functionality enabled by p-circuits is the concept of "invertible logic"[105] where for example, a Boolean circuit designed as a multiplier can be operated in reverse to factorize numbers, due to the reciprocal nature of p-circuit[65].

P-bits can mirror a special class of quantum circuits[106] by exploiting a well-known mapping between d-dimensional quantum systems and d+1-dimensional classical systems, a method often used in Quantum Monte Carlo calculations to simulate quantum systems in software. The basic idea is to represent a qubit network (d-dimensional) with a finite number (additional +1 dimension) of interacting replicas (d-dimensional) that are made from p-bits. Device level simulations show that spin-transfer-torque-based p-bits[66] interconnected with a resistive network can exactly reproduce the quantum correlations of the transverse Ising Hamiltonian, a system commonly used by quantum annealers[107]. It should be noted that even though the mapping between quantum and classical Hamiltonians is quite general, the mapped classical Hamiltonian can be efficiently simulated only for a subclass of quantum systems that does not suffer from the "sign" problem[108]. The sign problem arises when the quantum to classical mapping produces negative weights, making it exponentially hard to reduce errors in quantum Monte Carlo simulations. Whether a scaled hardware implementation of room temperature p-bits could be useful in emulating quantum systems with the sign problem in practical applications remains to be seen.

### d. Computing with nanomagnets

In the schemes described above, coupling between junctions is realized with CMOS circuits or by resistive crossbar arrays. Dipolar coupling between nanomagnets can also be exploited directly for computation based on energy minimization, decreasing the CMOS overhead of spintronic circuits. There are several demonstrations solving Ising Hamiltonians with nanomagnet arrays. Bhanja et al.[109] exploited the natural Hamiltonian describing the physical dipolar interaction between arrays of nanomagnets by mapping this interaction onto a quadratic optimization problem for computer vision applications. Debashis et al.[110] showed that small networks of nanomagnets interacting through dipolar fields can produce correlations corresponding to a Ising Hamiltonian. Nomura et al.[111] simulated a reservoir computer made of dipole coupled nanomagnets. In the future, such reconfigurable artificial spin glasses[112] could be interesting substrates for the implementation of scaled magnetic networks, enabling ultra-low power, high density co-processors by making use of the natural physics of nanomagnets.

### e. Computing with skyrmions

Towards even deeper miniaturization, Prychynenko et al.[82] proposed to use skyrmion assemblies (see Fig. 5b) as a fabric for reservoir computing. Here the reservoir is built out of a thin film of conducting material that hosts highly complex and self-organized patterns of magnetic skyrmions[83]. In this concept, the input signals are injected into the system through voltage patterns[84], ideally at randomly distributed contacts. The output signals are the different resistances measured between different contacts. Based on the interplay of spin-torques, pinning and magnetoresistive effects like the anisotropic magnetoresistance, an applied voltage across a certain magnetic texture leads to a complex current pattern. The underlying idea of this reservoir computing system is analogous to the water current pattern that arises in a riverbed filled with rocks, where water flow can induce changes in the arrangement of the rocks in the riverbed, in turn adjusting the current flow. In the magnetic case the current density relaxes on a much faster time scale than that of the magnetization dynamics (induced by the applied voltage patterns) allowing for self-consistent modelling.

The simulations in Ref.[82] show that single pinned skyrmions have non-linear I-V characteristics. The main effect of spin-torques on pinned skyrmions is their deformation. These in turn lead to a change in the current pattern and thus to a change in the measured resistance. For a single skyrmion, the effect of non-linearity is small, as it couples only to the size of the deformation. For larger effects, it appears beneficial to use more skyrmions, e.g. in the form of skyrmion assemblies. In addition to the basic requirements for any reservoir, basing a reservoir on complex structures that deform requires that the magnetic texture relaxes back to its original state when the voltage is turned off and that the system is stable under temperature fluctuations. In Ref.[84] the authors showed that skyrmion fabrics on top of a grain structure deform without significant displacement and are stable under thermal fluctuations, satisfying these additional requirements to operate as a reservoir.

Ref.[84] analyzes the response of the simulated system to different voltage patterns observing that the signal procession depends on the history of the reservoir, thereby showing a short-term memory. The complex magnetic response patterns serve as a high-dimensional nonlinear filtering of the input signals. Furthermore, the responses are the most non-linear close to the natural time scale of the system (nanoseconds in ferromagnetic systems). Simulations demonstrate simple pattern classification. This theoretical work shows that skyrmion fabrics are suitable for reservoir computing, providing a path to solve complex tasks using linear post-processing techniques based on nanostructures.

## V- Challenges for scaling up

The first experimental demonstrations of neuromorphic computing with small spintronics systems highlight the promise of this technology for future applications. However, deep networks implemented in software already comprise hundred millions of interconnected neurons and synapses for image recognition[113]. Several hurdles need to be overcome to scale up spintronic systems to sizes enabling useful pattern recognition. Some of these challenges are specific to spintronics, while others are shared by all technologies. In some cases, spintronics has advantages that could bring unique solutions for building large hardware neural networks.

### a. Adapting algorithms to spintronic hardware

Inference in hardware neural networks requires being able to read rapidly and precisely circuit outputs. A disadvantage of magnetic tunnel junctions compared to other memory technologies is their small resistance changes, which makes them difficult to read quickly[114], especially when they are multistate, with memristive-like behavior (their OFF/ON ratios are typically between one and three, while other resistive switching cells have ratios ranging from tens to millions). A way to circumvent this issue is to design circuits in which junctions do not need to be read individually. For example, the weighted sum of neuron voltages by the junction conductances, which is the important quantity for inference, can be read in the overall current flowing through the junctions connected in parallel, without any need to measure the resistance of each junction. However, this technique is limited to circuits of typically hundreds of junctions in parallel. Side stepping this issue requires complementing junctions with CMOS, either by connecting several small junction arrays with transistors, or by integrating a transistor below each junction. In both cases, these solutions limit the achievable density of the synaptic arrays.

Implementing neural networks that can be trained on-chip imposes additional constraints. Backpropagation algorithms[115] based on gradient descent require highly linear and symmetric weight variations. This is an issue for all emerging memories and for most memristor types which tend to have highly nonlinear asymmetric responses[116]. One approach to achieve this linearity with spintronic memristors is to tune the materials and mechanisms underlying resistive variations, by considerably shrinking the size of domain walls or skyrmions down to a few nanometers[117] to decrease granularity. In parallel, within the artificial intelligence community, there are considerable efforts to develop algorithms based on weights with reduced precision. For example, complex neural networks have been trained with only eight bits per synapse for the weights[118]. For inference, extremely reduced precision may be used: in 2016, it was shown that for many situations, binary weights are appropriate, which is well adapted to encode in magnetic tunnel junctions[119,120]. Finally, the stochasticity inherent to magnetic systems can be a problem but one that can possibly be turned to an advantage for accelerating training. Continuous training of neural circuits during the inference phase offers other potential advantages, when coupled with the large cyclability of spintronics systems, particularly magnetic tunnel junctions[98].

**b. Low energy**

Neuromorphic systems are most useful if they use less energy than traditional approaches for particular computational tasks. At the system level, it is important to reduce CMOS overhead in spintronic circuits, by taking advantage of physical effects to achieve functions that CMOS does not do well. It is also important to keep the energy consumption low for individual devices. As in most non-volatile memories, the write energy of magnetic tunnel junctions is higher than their read energy, and should be therefore be considered carefully during learning. The write energy consumption of magnetic memory cells today is of the order of a few hundred femtojoules per bit, lower than phase change memories and comparable to redox memories[6]. To decrease this energy consumption further, three options are available. The first is to improve spin-torque efficiency, for example through the use of spin-orbit torques provided by topological insulators[121]. The second is to speed up devices, for example by combining ultrafast demagnetizing process with parallel optical writing[122,123] and reading or using antiferromagnets to generate magnetization dynamics in the terahertz range[124,125]. The third, already mentioned in this review, is to decrease the size of the devices to the point that thermal fluctuations help electric currents drive magnetization dynamics, for example in the superparamagnetic limit[126].

**c. Interconnection**

A major challenge for neuromorphic hardware is to reach a high degree of interconnection between neurons. There are from 10 to 1000 synapses per neuron in typical algorithms today, in contrast to the 10,000 synapses per neuron in the cortex. There is no good solution today to reach such degree of interconnection while keeping the related power consumption low. Spintronics offers interesting opportunities in this domain. Spintronic systems are made of multilayer systems that naturally stack in three dimensions[127]. It is therefore possible to envision building three-dimensional spintronics neuromorphic systems exploiting solitons such as domain walls, skyrmions or magnons for vertical and horizontal communication. Communication through optical waves, or microwave signals emitted by spin-torque nano-oscillators is also potentially useful for this purpose, but amplification through external circuitry could be required to achieve high fan-out. Progress in spintronics materials and nanodevices now offers the possibility of building complex three-dimensional computing systems[74].

In summary, based on the basic principles of how brains compute, spintronics could help realize artificial intelligence in at least two ways. First, it allows enmeshing computation and memory at a very local level. Second, it permits exploiting rich multiphysics as a source of computational power. The recent experimental progress achieved by several groups delivers the first proofs of concept and pushes toward the development of large scale brain-inspired spintronic systems.

**Acknowledgements**

Work by M.D.S. was supported by the U.S. Department of Energy (DOE), Office of Science, Office of Basic Energy Sciences (BES), Materials Sciences and Engineering Division under Award DE-SC0019273. SF is funded by JSPS Grant-in-Aid 18KK0143 and JST-OPERA. KES is funded by the German Research Foundation (DFG) under the Project No. EV 196/2-1 and acknowledges support through the Emergent AI Center, funded by the Carl-Zeiss-Stiftung. Work by J.G. was supported by the European Research Council ERC under Grant bioSPINspired 682955. Work by D.Q. was supported by the European Research Council grant NANOINFER (reference: 715872). S.F. acknowledges discussion with Hideo Ohno. K.E.S. acknowledges discussions with Daniele Pinna.

**Data availability**

The datasets generated and analysed during this study are available from the corresponding authors on reasonable request.

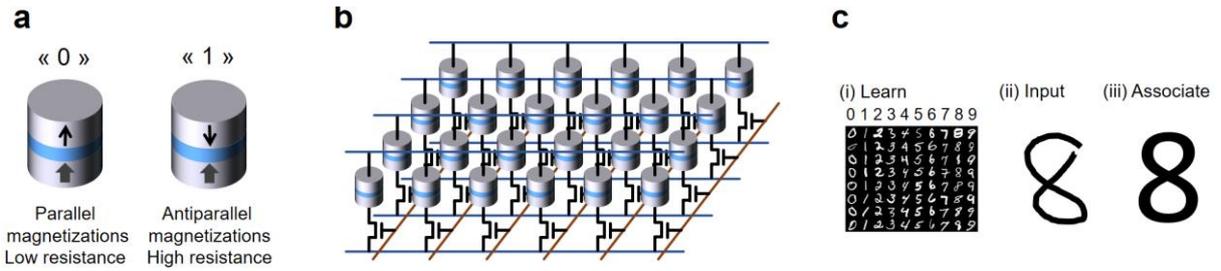

Fig. 1. (a) Magnetic tunnel junctions for memory applications. A magnetic junction consists of two ferromagnetic layers (gray) separated by an insulating layer (blue) with the magnetization of one layer fixed and that of the other either parallel (low resistance) or antiparallel (high resistance) to it. (b) Cross-bar array of magnetic tunnel junctions for high density storage (Magnetic Random Access Memory). The resistance of a particular tunnel junction is measured by activating the appropriate word line (red) allowing conduction between the bottom bit line and the top sense line (both blue). The alignment of the magnetization can be switched by passing sufficient currents through the device. (c) Associative memory. (i) Handwritten digits from the MNIST dataset used for training the associative memory. (ii) Sample test input after training. (iii) Output of trained network from the test input showing successful association.

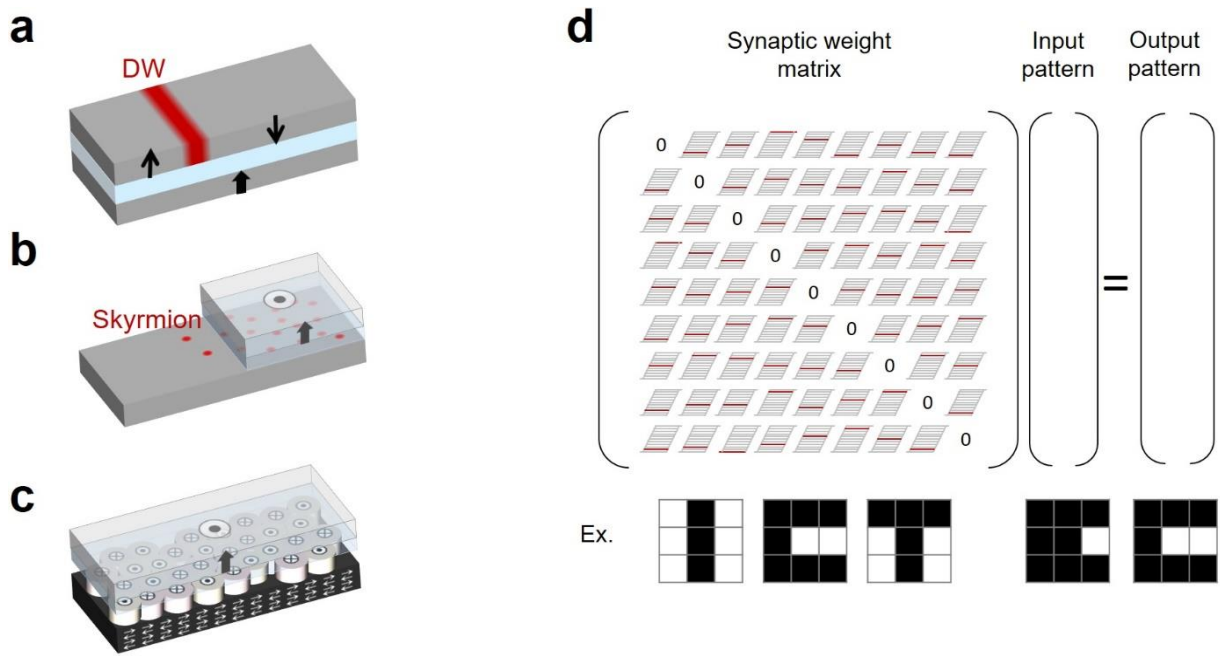

Fig. 2. Spintronic based memristors. (a) Domain wall memristor. The resistance of the magnetic tunnel junction depends on the location of the domain wall changing the relative area of the high resistance antiparallel configuration and the low resistance parallel configuration. (b) Skyrmion based memristor. the resistance of the device depends on the number of skyrmions under the fixed layer. (c) Fine-magnetic-domain tunneling memristor. In a tunnel junction coupled to a polycrystalline antiferromagnet, the variation of switching properties from domain to domain allows the domains to reverse independently and under different conditions. The resistance of the device then depends on the fraction of domains with magnetizations aligned with the uniformly magnetized fixed layer. (d) Spintronic associative memory. The value of each off-diagonal matrix element is stored in the configuration of the memristor schematically illustrated by the different levels in the matrix. These levels are trained so that when the matrix multiplies an input, the result is the closest element of the training set. The multiplication is carried out by applying voltages that corresponding to the input and measuring the output current through the appropriate memristors.

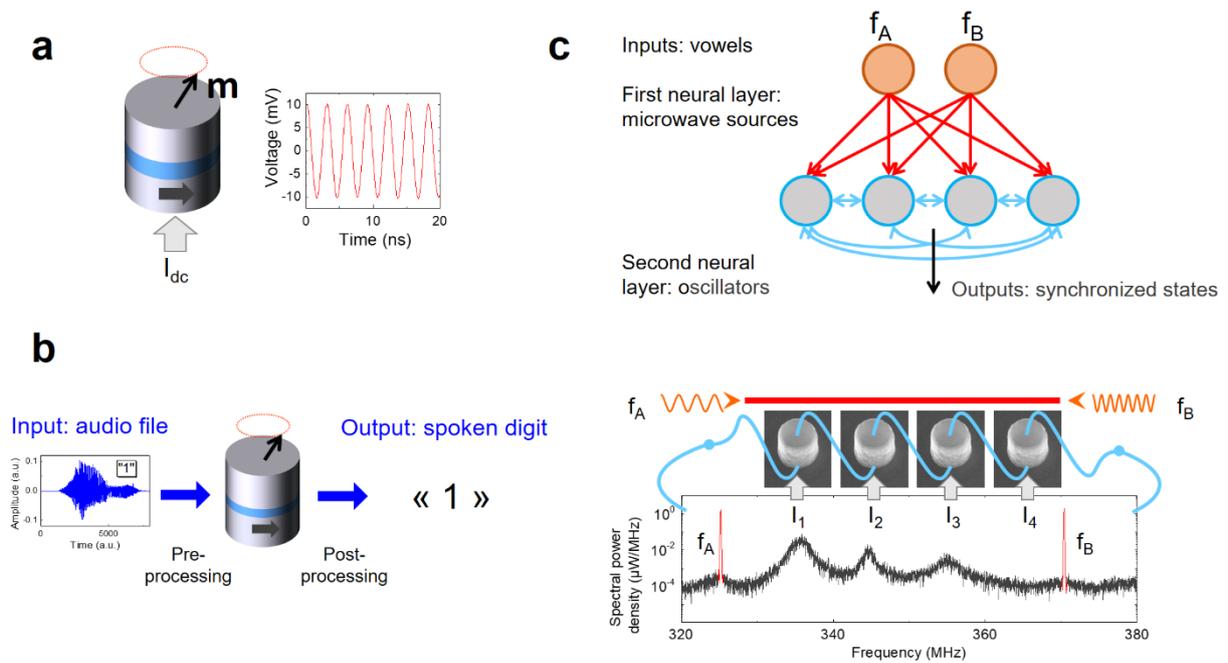

Fig. 3. Neuromorphic computing with Spin Torque nano-oscillators. (a) Schematic spin torque nano-oscillator. When designed appropriately, the free layer magnetization of a tunnel junction precesses when a dc current is passed through it. Because of the oscillating magnetoresistance, a fixed input current gives an oscillating voltage across the junction. (b) Reservoir computing with a spin torque nano-oscillator. Using time multiplexing in pre- and post-processing, a single spin torque nano-oscillator gives state of the art performance as a reservoir in a reservoir computing scheme. (c) Schematic use of coupled nano-oscillators for vowel recognition. The input is represented by the frequencies of two microwaves applied through a stripline to the oscillators. The natural frequencies of the oscillators are tuned by dc bias currents through the devise. These can be tuned so that the synchronization pattern between the oscillators corresponds to the desired output.

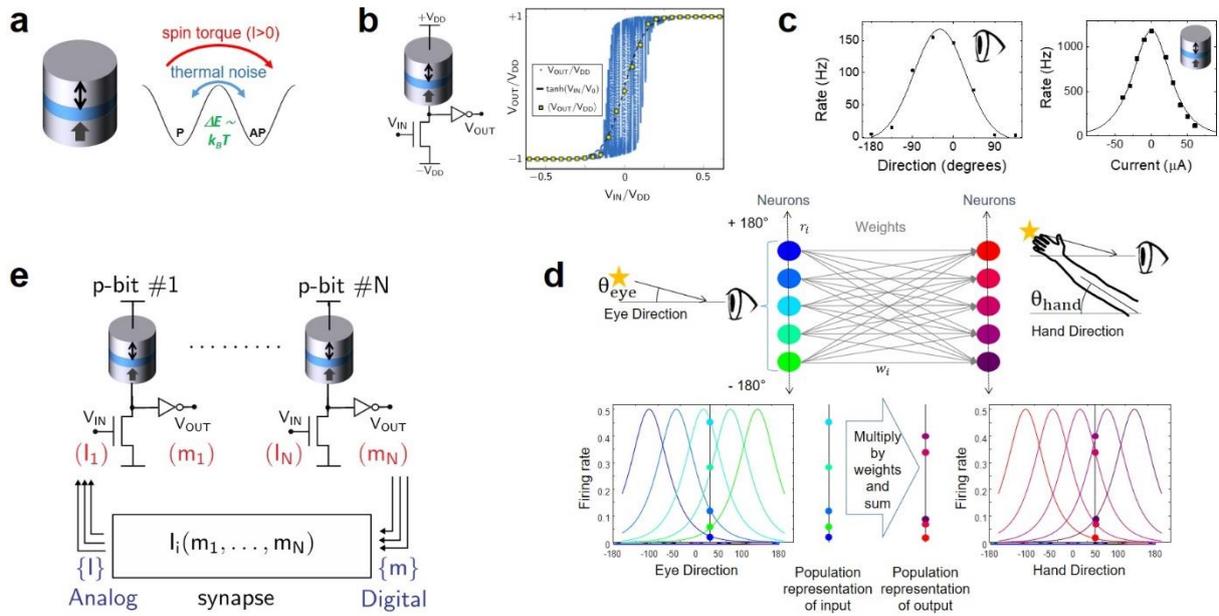

Fig. 4. Computing with stochastic magnetic tunnel junctions. (a) Stochastic magnetic tunnel junctions. Thermal fluctuations cause stochastic transitions between the low resistance parallel and high resistance antiparallel states. Spin torque can bias the fluctuations favoring one configuration over the other depending on the sign of the current. (b) P-bit. Varying the bias on the input transistor controls the current through the magnetic tunnel junction controlling the fraction of time the tunnel junction spends in each magnetic state. The simulated average output voltage agrees very well with the expression $\langle V_{\text{out}} \rangle = V_{\text{DD}} \tanh(V_{\text{in}}/V_0)$, where $V_0$ depends on the temperature. (c) Tuning curves. Sensory neurons in the eye fire with rates that are highest when the eye is oriented in a particular direction. Similarly, the stochastic transition rates (rather than the average times spent in each state as in (b)) in magnetic tunnel junctions decrease with current in both directions. (d) Population-coding-based computation. In population coding, an input value is represented by the firing rates of a set of neurons each of which is tuned to be most sensitive to different input values. Non-linear computations can be performed on sets by multiplying the rates by synaptic weights giving a population-based representation of the output value. Here the input is the angle the eye observes an object at and the output is the angle the arm needs to make to grasp the object. The two angles are around different points and are non-linearly related. (e) Computing with p-bits. The analogue input voltage to each tunnel junction produces a fluctuating (digital) output voltage whose average depends non-linearly on the input as in (b). The fluctuating output voltages are combined with appropriate synaptic weights to produce the analogue current inputs to the p-bits.

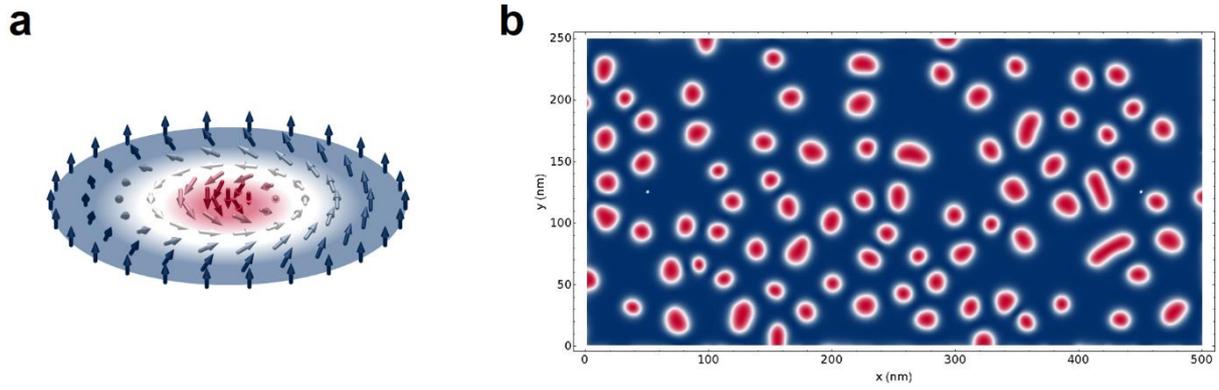

Fig. 5. (a) Schematic skyrmion structure. The magnetization direction of a single skyrmion is schematically given both by the directions of the arrows and the color coding, ranging from blue for magnetization up, through white for in-plane magnetization directions, to red for magnetization down. (b) Simulated skyrmion assembly. A reservoir computing scheme based on skyrmions in a random potential makes use of the distortions of the assembly due to current flow to provide the necessary non-linearity and memory.